# Satio Hayakawa and dawn of high-energy astrophysics in Japan

Jun Nishimura

*ISAS/JAXA, Sagamihara, Kanagawa, Japan*

(Abstract) Gamma ray astrophysics is now one of the most exciting fields in the space physics, in which the Fermi satellite has been playing an important role in exploring new phenomena and findings. Needless to say, great strides were also made recently in higher energy region by the ground based IACT (Imaging Air Cherenkov Telescopes) of Veritas, HESS and Magic. The closely related fields of the gamma-ray astrophysics, X-ray astrophysics as well as the direct observations of Cosmic-ray particles have given us exciting information on the violent phenomena occurring in the stars, our Galaxy, and other galaxies as well as Intergalactic space.

I am most pleased to talk in this 5th Fermi symposium at Nagoya University, where Hayakawa spent his most active time as a pioneer and an outstanding leader, promoting and organizing the young physicists in a wide range of physics topics, particularly in the field of high-energy astrophysics including Infrared astrophysics in the space.

## 1.

As in the case of other countries, in our country, cosmic-ray physicists first promoted high-energy astrophysics. Hayakawa started cosmic-ray studies under S.Tomonaga, in the field of the high-energy particle physics. Soon after, around 1950s, his interests move to the cosmic–ray studies as an approach to high-energy astrophysics. He anticipated that it would become a central topic in near future when not so many scientists had paid attentions to this field yet.

As early as 1948, Feinberg and Primakoff [1] discussed the energy loss of cosmic-ray electrons by the Inverse Compton process between cosmic-ray electrons and star lights. Some of the photons boosted by high-energy cosmic-ray electrons in this process are emitted as gamma rays in the space. However, the flux of gamma rays was estimated to be very small, and detecting them was thought to be difficult. This may be the first prediction of gamma rays from space, outside of gamma rays from the Sun.

Four years later, in 1952, Hayakawa pointed out the significance of the gamma-ray astrophysics predicting galactic diffuse gamma rays from the decay of $\pi^o$ produced in the collisions of cosmic rays with interstellar matters [2]. Since his flux estimate was also small, most cosmic-ray physicists were reluctant to attempt experiments, because it would be extremely difficult to detect the gamma rays under the strong background of cosmic rays. In the same year, Hutchinson also estimated the relative intensity of bremsstrahlung gamma rays by the collisions of high-energy cosmic-ray electrons and interstellar matter [3].

Six years after Hayakawa's prediction, P. Morrison advocated the importance of the gamma rays in the high-energy astrophysics, and predicted the most optimistic estimates so far of gamma-ray flux from the space [4].

This prediction encouraged the cosmic-ray physicists, since it might be easier to detect the gamma rays with rather simple detectors. His prediction was so optimistic, and in some cases it was several orders of magnitude higher than what we observe in recent measurements.

Several balloon experiments tries to detect gamma rays from the space, but with a disappointing lack of success until the gamma-ray satellite OSO-3 first succeeded in observing significant indication of gamma from the Galactic disc [5]. The results of OSO-3 almost agreed with predictions by Hayakawa.

After OSO-3, Gamma-ray Satellites SAS-2, COS-B, CGRO (Compton Gamma ray Observatory), and Integral were launched, and today the Fermi Gamma-ray Satellite has been in operation since 2008. Now gamma-ray astrophysics is one of the most important ways to explore the violent phenomena in the Universe.

The prediction of fluxes in X-ray astrophysics came almost ten years after that of Gamma-ray astrophysics, but X-ray stars were successfully detected in 1962, almost five years before the first significant detection of gamma rays by OSO-3. X-ray astrophysics is closely related to gamma-ray astrophysics, and our understanding of the high-energy phenomena in the space are naturally performed in connection with the results of gamma-ray astrophysics.

Hayakawa also presented several important arguments in high energy astrophysics including:
- Super Nova origin of Cosmic rays
- Long lived Radio Isotope $Be^{10}$ as a spallation fragment from primary cosmic rays in Galactic Space
- High Energy primary Electron, and others

Some details of these topics are in the following sections.

## 2. Birth of Cosmic-ray Studies in our country

Around 1930, several laboratories had started cosmic-ray research in our country. Among those the Nishina laboratory in Riken, was the largest scale efforts, and conducted most comprehensive researches in this field.

Y.Nishina returned to Japan in 1928, after spending several years studying the modern physics in Europe. Nishina is known as one of the authors of the paper of presenting the Klein-Nishina formula of Compton scattering. This work was performed in Bohr Institute before he left Copenhagen for Japan. He believed it was the most important to extend the Modern Physics in our country, and he asked to Riken to invite the distinguished scientists to introduce Modern Physics to Japan. Heisenberg and Dirac were invited in 1929, and they gave a series of lectures at the University of Tokyo. Nishina himself also lectured on the Modern Physics in a few universities. Yukawa and Tomonaga were graduate students in those days, and they were greatly stimulated to study this field by the lectures.

Nishina laboratory was founded in Riken in 1931. He created four groups in his laboratory; i.e.,

● Theory
● Cosmic rays
● Nuclear Physics by constructing Cyclotron on the same scale as the largest one in Berkley, US.
●Radio biology.

One of the achievements of cosmic ray research in this laboratory was the identification of mesons in cosmic rays, by constructing magnetic cloud chamber of 40 cm dia. with magnetic field of 1.7 T. In 1937, Y.Nishina, M.Takeuchi and T.Ichimiya succeeded to observed the Muon track in their chamber [6], at almost simultaneously with similar works by Neddermeyer–Anderson [7] and Street-Stevens [8].

Nishina-Takeuchi-Ichimiya identified the mass of a meson from the track in their magnetic Cloud Chamber as 223 ± 36me. This was the most accurate measurements in those days, and was within a range of the most recent values of 206.768…me. When they found this Muon track, Nishina immediately contacted Yukawa, informing him that the track is most likely the meson Yukawa has predicted. The arguments that who found the Muons first are presented in reference [9].

Parallel to the research of this magnetic Cloud Chamber, the Nishina laboratory observed the cosmic-ray intensity deep underground at 1400m.w.e. to 3000m.w.e. during 1939-1944 at Shimizu Tunnel, which is, locates almost 150km North-Northwest from Tokyo [10]. The observed intensity at 3000m.w.e. was the deepest point data before the observation by Bollinger in US was established in 1951 [11].

Plans were made for continuous observations of cosmic-ray flux at five different latitudes of Sakhalin, Hokkaido, Tokyo, Taiwan and Palau, and construction of five stable ionization chambers named as Nishina-Type was set in motion in 1935. However all of those chambers were kept in Tokyo because of the War II, but successful to observe the first Forbush increase from the Solar flare in 1942 [12]. Latitude effect surveys and balloon observations were also performed.

During the World War II, the experimental works were suppressed, however, significant progress were continued in the theoretical physics, with semi-regular meeting held relating to meson theory. It is to be noted that the two meson theory was proposed by Sakata, Inoue, Tanikawa already in 1942, in advance to Marshak and Bethe (1947) to resolved the conflicts between lifetime and interaction cross-sections observed in cosmic rays and those of theoretical prediction. In relation to this two-meson theory Taketani also proposed that the gamma rays from the decay of the neutral mesons are the main source of the soft components of cosmic rays in the atmosphere in 1942. Some details of those of the works in the Nishina laboratory are found in the reference [9].

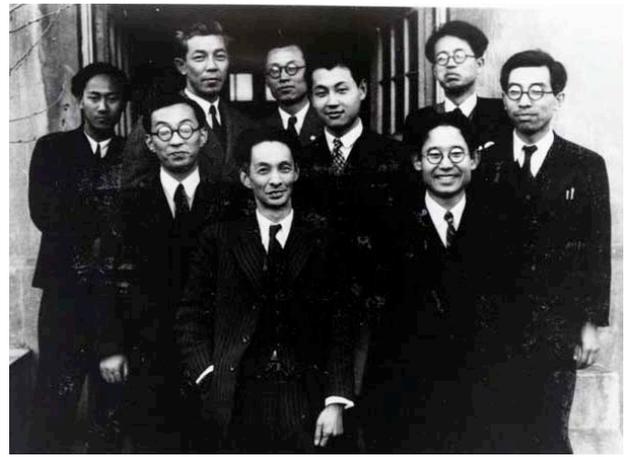

Figure 1: Tomonaga with his Colleagues and Students Hayakawa is right behind Tomonaga. Around1950.
Left to right: Front: S.Sakata, S.Tomonaga, M.Taketani.
Middle: K.Baba, S. Hayakawa, T. Miyajima.
Back: O.Minakawa, T. Kinoista  J. Koba.
(From: Tomonaga Memorial Room, University of Tsukuba)

After the War-II, Hayakawa studies cosmic rays under S.Tomonaga in relating to the works of the Nishina laboratory, as the field of the high-energy physics. He provided the analysis of depth and intensity relation observed deep underground in Shimizu Tunnel. He showed that the intensity depth curve bending from the power spectrum can be well explained as the effect of π-μ decay life time including energy losses by the processes of Radiation, Photo-nuclear reaction and Direct pair creations by muons in 1949 [13]. K. Greisen published the same concept on the effect of the π-μ decay to the depth intensity curve independently in US at almost the same time  [14].

## 3. Gamma-Ray Astrophysics

### 3.1. Gamma-ray Astrophysics predicted by Hayakawa

Hayakawa first concentrated his effort on cosmic rays as an approach to the field of particle physics, but around the 1950s, his interests also extended to cosmic-ray research as high energy astrophysics. He found also it may be favourable given the situation in our country, since cosmic-ray research in high energy astrophysics does not quite require the most recent accelerator results as does cosmic-ray research as the particle physics.
Our country is remote from where the work in high-energy accelerator physics was centered.

He made significant contributions himself, and stimulated the young scientists to work in this field.
In his paper on:

*"Propagation of the Cosmic Radiation through Interstellar Space,*
*S. Hayakawa, 1952, Prog. Theor. Phys.8, p571",*

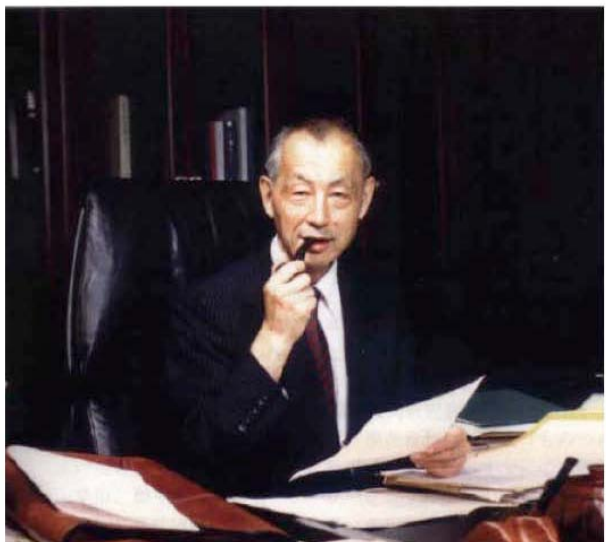

**Figure 2:  S. Hayakawa**
He performed comprehensive works on Cosmic rays and high-energy Astrophysics,  and predicted the importance of Gamma-ray astrophysics through $\pi^0$ decay in 1952.

he discussed how much interstellar matter is traversed by cosmic rays during he transport from the sources to the Earth, referring to the observed data on heavy primaries in cosmic rays by two groups of the Bradt – Peters (1950), and Dainton-Fowler-Kent. (1951).

In his paper, he also mentioned that gamma-ray emission from the $\pi^0$ mesons produced in collisions of cosmic rays and the interstellar medium during the propagation of cosmic rays, is such that:

" *In this passage through this thickness secondary particles are scarcely produced except photons which are due to the decay of neutral pions. The intensity of the secondary photons are estimated as about 0.1% of the total intensity at the geomagnetic latitude 55°, but as nearly 1.5% at the equator*".

This means, Hayakawa predicted the gamma ray flux of
$$\sim 2 \times 10^{-4}/cm^2 s.sr.$$
which almost agree with recently accepted data.

The concept was accepted that it is important for gamma rays from space to be observed, since the gamma-ray flux is proportional to the amount of matter in the line of sight that is,
*(Cosmic rays density) times (Density of Interstellar medium).*

Thus observation of the gamma ray flux bring us important information on the density of Cosmic rays and Interstellar medium in space, which would be difficult to obtain in otherwise.

However the flux was so faint, almost all cosmic ray scientists were reluctant to attempt experiments with the detector technologies of the time, since they thought that the extraction of the gamma ray flux is very difficult given the strong background of cosmic rays.

## 3.2. Gamma-ray Flux Predicted by  P.Morrison [4]

Six year after the perdition by Hayakawa, P Morrison advocated the importance of Gamma ray astrophysics in 1958. His main argument is that the astrophysics was developed in the past by observing visible light and radio wave, but those photons were descendants of the gamma rays produced from the high-energy phenomena in the stars and Galaxies. In this respect, it is important to observe directly the gamma rays from the source to understand what are happening in the space.    Instead of diffuse gamma rays estimated by Hayakawa, he focused to the point sources of gamma rays of the Active stars  and Galaxies.

He first discussed on the processes of gamma-ray production:
Synchrotron, Bremsstrahlung, Nuclear gamma rays, $\pi^0$-decay, Matter and Antimatter annihilations.

In the case of the Radio luminous colliding galaxies of Cyg-A, he estimated the gamma ray flux by assuming the energy source of galaxy is due to the matter – antimatter annihilation. His estimated flux of Gamma rays of Cyg-A was
$$0.1\text{-}1.0/cm^2 s$$
in the range of a few MeV to a few hundred MeV. This is several orders higher than Hayakawa's estimate for the flux of diffuse gamma rays.

Then Morrison proposed we could observe the gamma rays rather easily, if we point the detectors to the source.

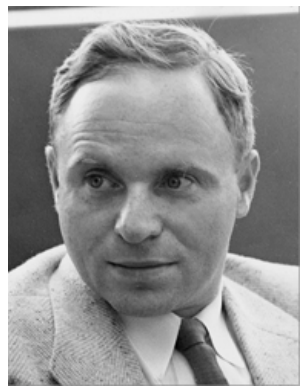

**Figure 3 :  P. Morrison**
He advocated the importance of the Gamma–ray Astrophysics particularly on the point sources in 1958.

In his paper [4]:
     "*On Gamma-ray Astrophysics*
          *P. Morrison*
    *1958, IL. Nuovo Cimento VII, No.6, 858*",

he mentioned that  *:*
"*Flights of several hr.'s duration are adequate, and the altitude required are  not extreme. Telemetering of data,*

*or even recovery of the apparatus with stored data. Reasonable angular discrimination can perhaps be obtained in the low energy region at least using lead collimation, should balloon loads permit. Otherwise, the use of scintillation counters, possibly taking advantage of coincidences with Compton scattered photons to help define angles, seems capable of adequate energy and angular discrimination below 1 or 2 MeV. The dominance of pair-production makes counting techniques even more satisfactory, and angular discrimination easier, in the energy range from 10MeV to a few hundred MeV. Here emulsion might be of value."*

This statement encouraged many scientists to carry the balloon observations, but they were unsuccessful till the significant observation was made by OSO-3 [5] in late 1960s, almost ten years after the prediction by Morrison.

The importance of Gamma-ray astrophysics, however, has been well recognized by those papers of Hayakawa [2], Morrison [4], together with the as-yet unsuccessful experiments to detect the gamma rays.

In fact, I remember his speech at the dinner party of ICRR (International Conference of Cosmic-ray Conference), in Kyoto in 1961, C. F. Powell, the Nobel Laureate in 1950, said :

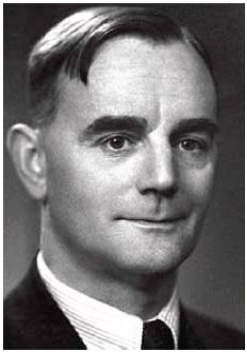

Figure 4.  C.F. Powell
Nobel Laureate for identifying Pions and Muons using Nuclear Emulsions.  He served as a Chairman of Cosmic Ray commission of IUPAP.

*"In the near future, we cosmic ray physicists shall tell to the Astronomers !*
*How much interstellar matter there is, and how it is distributed in our Galaxy !"*

### 3.3. Short Summary

Explore-XI was the first gamma-ray satellite and detected 31 gamma rays during 7month, but later they found it was suffered by heavily backgrounds [15]. The same group improved the detectors and put then on board the OSO-3. OSO-3 detected high-energy gamma rays (>50MeV) from the Galactic plane for the first time in 1967-68 fifteen years after Hayakawa's prediction [5].

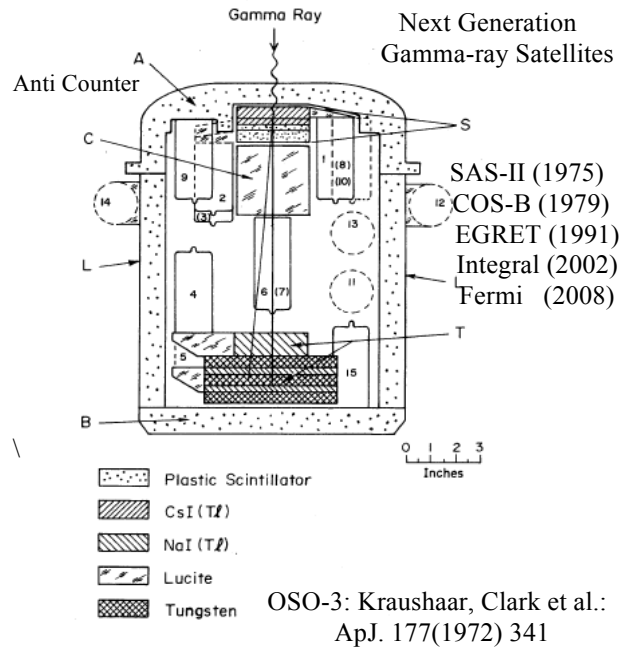

OSO-3: Kraushaar, Clark et al.: ApJ. 177(1972) 341

Figure 5: Gamma-ray satellite OSO-3 [5], and next generation satellites.

They equipped CsI and NaI scintillators arranged as a "phoswich" detector, combining Cerenkov counters inside shielding counters made of plastic scintillators. A total number of 621 gamma rays were observed by this satellite during 16 month in approximate   agreement with Hayakawa's prediction.

Following to OSO-3, Satellite SAS-2 COS-B, EGRET, Integral were launched, and the Fermi Satellite are now in work since 2008.  SAS-2, Cos-B and EGRET equipped the spark chamber as the imaging detectors to identify the pair electrons from gamma rays without ambiguity from the background tracks. These satellites with spark chambers may be called the second generation of the gamma ray satellites. Integral and Fermi are considered as the third generation of the gamma ray satellites equipped with sophisticated solid state detectors and electronics instead of spark chamber, and thus can analyse the large amounts of data with high statistical accuracy. In particular Fermi satellite can observe gamma rays of energy range extended up to several hundred GeV with high accuracies. Then the gamma-ray astrophysics developed to one of the most significant field to explore the violent phenomena in the Galaxy, and in the Active Galaxies.

When we recall the beginning of the gamma-ray astrophysics, we found:
Hayakawa's prediction was relatively accurate, predicting so faint flux. Then, cosmic-ray physicists reluctant to attempt the experiments.
On the other hand, Morrison's Prediction was optimistic, and encouraged the physicists to carry out gamma–ray detection experiment.  His optimistic estimation surely opens the door of the gamma ray astrophysics.

We found the Irony what happed in this history that:

" *Accurate expectations do not always help to open the*
   *door of new field,*
                        *but*
   *Optimistic and even somewhat erroneous expectation*
   *promoted to start the Gamma–ray Astrophysics.* "

## 4. X-ray Astrophysics

In contrast to gamma-ray Astronomy, X-rays from space got little attention in the 1950s.

B. Rossi at MIT, had discussions to see the possible observation of X rays from Celestial bodies except to the Sun early in 1960. Hayakawa also joined the meeting. No promising objects for X–ray emission sources were proposed, since the role of compact stars and the extremely high magnetic field of the neutron stars were not yet well understood at the time. The only possibility discussed was faint fluorescent X rays from the lunar surface produced by solar x-rays or cosmic rays.

However, Rossi commented:
   *"The Nature is more imaginative in many case than we suspect!! ,"*
and requested a sounding rocket mission with three Geiger counters on board, an excellent decision.

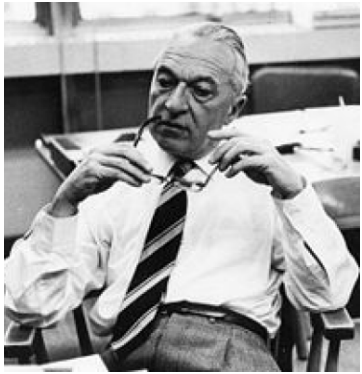

Figure 6 :        B. Rossi
Rossi is a pioneer of on Cosmic ray research since early 1930s, and also the originator of X-ray astrophysics with his MIT colleagues in 1962 [16].

He and his colleague found an extremely strong X-ray flux from the direction of Scorpio X-1 [16].   This is the beginning of X-ray astrophysics.

It is interesting to note that the prediction of X-ray astrophysics came 10 years later than Gamma-ray astrophysics, but the first successful observation was almost 5 years earlier than that of gamma rays.

As to the start of X-ray astrophysics in our country, it was important that M.Oda was asked to join to the MIT group by Rossi, for the early development of the X-ray astrophysics. The reason he was asked to join was that Oda was temporarily in the laboratory of Rossi early 1950's to work on the Extensive Air Showers.

Oda invented the modulation collimator (Fig. 7) during his stay at MIT, and successfully located the position of the optical counterpart of SCOX-1.

In 1965, when the Institute of Space and Aeronautical Science, the Collaborative Institute of Space Science in our country, was founded in the University of Tokyo, he came back to the Institute and spent much effort to develop the X-rays astronomy in our country.

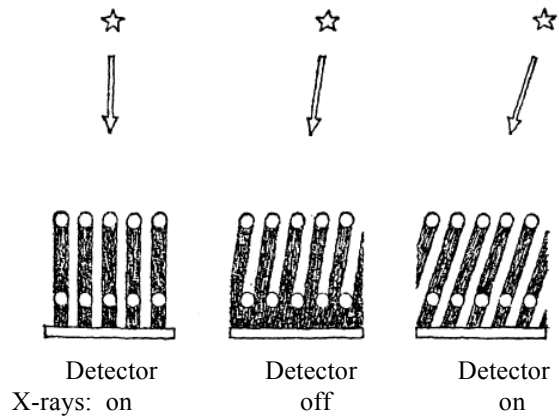

Figure 7 :  Modulation Collimator invented by Oda . By observing the time modulation of the point source, he could locate the X-ray sources with wide field view of detectors.

M. Oda and S. Miyamoto, staff of Oda's laboratory, S. Hayakawa himself and Y. Tanaka, staff of his laboratory had push forward this field in our country.

Unlike Gamma-ray astrophysics, X-rays were detectable by simple detectors, because of their high intensity. In this respect, X-ray astrophysics has been attractive to scientists in our country, where our space facilities had only small-payload launching capability until recently.   One of the achievements with modulation collimators in early days by balloon observations are shown Fig. 8 and 9, which was to locate the precise position of CygX-1 [17].

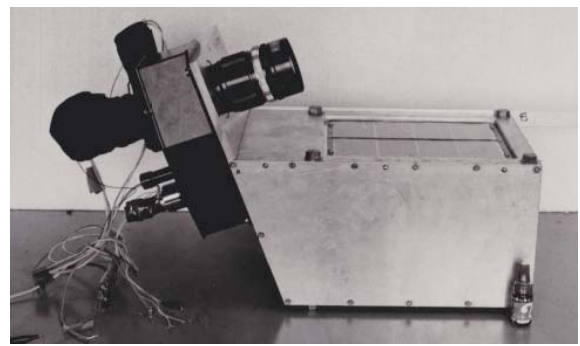

Figure 8. Balloon Borne Detector to locate the CygX-1 with Modulation Collimator.
The Right hand side in Figure 8 is the detector with Modulation collimators to observe the location of CygX-1. Optical Telescope on the left hand side observes the location of the known stars to identify the absolute direction of this detectors [17].

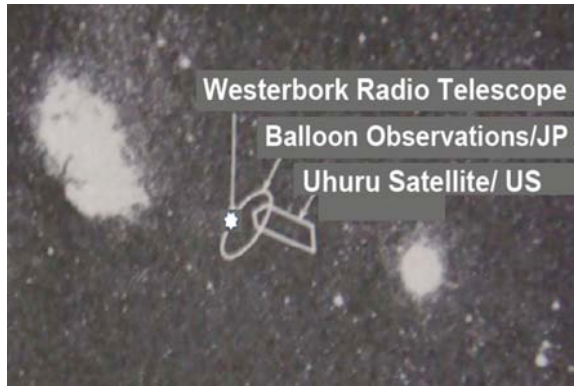

Figure 9: Location of CygX-1 observed by Uhuru Satellite in US and Balloon observations in Japan with the detector shown in Fig. 8.

The location predicted by each group agrees within an error box of each group of several arc min. Soon after Westerberg Radio Telescope find a radio source in these area and Pinpointed the location.

At almost the same time, similar work with Uhuru, the first scientific satellite for X-rays, was performed in the US [18], and the results agree with each other as shown in Fig.9. The X-ray source location was examined by the radio telescope at Westerbork, and a variable radio source was found. Then the optical counterpart was identified, and it was found that the source is associated with a heavy non-visible star of almost 15 times of mass of the sun. Thus, CygX-1 was presumed to be the first candidate of Black hole.

The first Japanese X-ray satellite, Hakucho, was launched in 1979, and next X-ray satellites followed at intervals of a few years.

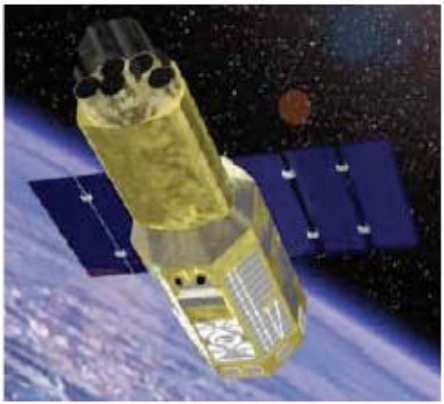

Figure 10: Suzaku
High sensitivity, Soft X-ray Imaging Spectroscopy and Wideband of soft to hard X-ray Spectroscopy.
About 1.7 tons weight, in orbit since 2005.

Although Japanese satellites were small compared to the satellites of other countries in those days, we provided the important advantage of quick response to new findings, to successfully develop X-ray astrophysics in our country. In recent years, however, observations are required more precise, and the satellite required becomes heavy. We now have Suzaku satellite in operation, and Astro-H is to be launched within a few years. The on board instruments are getting more sophisticated as shown in Fig. 10 and 11.

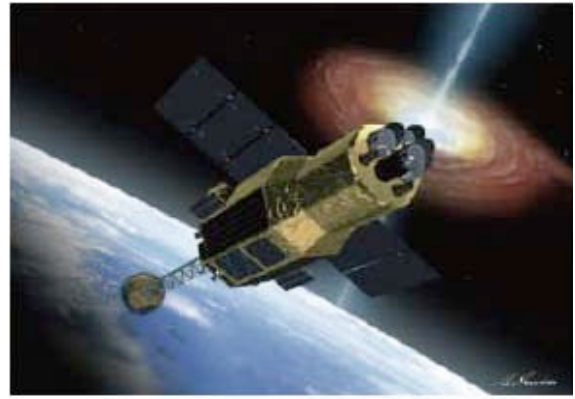

Figure 11. Astro-H
High Resolution Soft X-ray Spectroscopy, and High-sensitivity hard X-ray imaging spectroscopy.
About 2.7 tons weight to be launched in 2015 to early 2016.

# 4. Cosmic rays

Origin of Cosmic rays, transportations from source to the Earth and the composition of Cosmic rays are closely related to the high-energy phenomena occurring in the Galaxy. In this respect, Hayakawa made several significant contributions. Some of those are:

## 4.1 Supernova Origin of Cosmic Rays (1956)

W. Baade and F. Zwickey first proposed the model of "Super Nova Origin of Cosmic rays" based on the large energy release of Super Nova explosion (1934) [19]. Later I.S. Shcklovsky and V.L. Ginzburg extended this concept on the bases of the strong radio wave and visible lights from Crab nebula are assumed as the synchrotron radiation by the high energy electrons accelerated in the supernova, and predicted the light should be polarized in 1950s [20], [21]. This model was supported by the observations of polarized of light by J.H. Oort and T.H. Walraven in 1956 [22].

Hayakawa approached the problems in a different way by focusing the relative abundance of the composition of cosmic rays at the source after correcting the fragmentations of heavy elements during the transportation. He presented the model of "Supper-Nova Origin of Cosmic rays" based on the relative overabundance of heavy nuclei in cosmic rays. A super nova is the last stage in the evolution of a massive star, when the relative abundance of heavy elements is large. This argument was accepted to support the model of super nova origin of comic rays when it was published. [23].

Stimulated by his work, more detailed arguments have been developed later and discussions are now still continued to identify the sources taking account of the compositions of cosmic rays and the possible sources.

## 4.1. Be$^{10}$ as to the measure of confinement time of Cosmic rays

During the transportation of cosmic rays from the source to the Earth in the Galaxy, he mentioned the importance of the long lived radioactive nuclei such as Be$^{10}$ ($\tau \sim 1.5*10^6$yr) spallation products of the collisions between cosmic rays and interstellar gas. The fraction of survived Be$^{10}$ gives us the information how long cosmic rays were confined in our Galaxy. Thus the fraction of Survived Be$^{10}$ constrains the amount of cosmic rays required to be produced per unit time in our Galaxy [24]. i.e., the acceleration efficiency of cosmic rays.

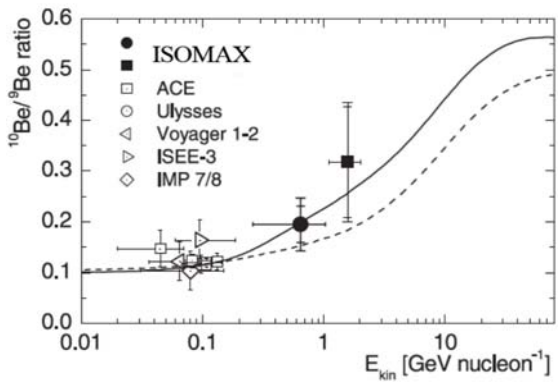

Figure 12: Be$^{10}$/Be$^9$ , From [26] of ISOMAX. (2004). Most recent data published by Pamela group (2013) [27] around 1GeV/Nucleon are not included here, which are consistent with that of ISOMAX [26]

The small flux of Be$^{10}$ is difficult to detect, and was first observed around 0.1GeV/nucleon by Garcia Munos et al. using the IMP 7 and 8 satellites, in 1977 [25]. Around 1GeV, ISOMAX and Pamela with magnet spectrometer succeeded in observing the Be$^{10}$ [26], [27]. Those results are shown in Fig. 12, which indicate the confinement time of the cosmic rays is about 10$^7$years around 1GeV. More detailed observations will be made in the near future, which may allow a more precise estimate of the confinement time of cosmic rays in our Galaxy.

## 4.2. Cosmic-ray Electrons

Unlike other cosmic-ray components, primary cosmic-ray electrons loss their energy primarily by Synchrotron and Inverse Compton processes during transport from the source to the earth. Since these energy losses are approximately proportional to the square of the electron energy, the spectrum of the electrons adds an interesting feature, particularly at higher energies. Positrons in cosmic rays are naturally produced by the decay of muons produced in collisions between cosmic rays and interstellar medium, and in fact the observed positron intensity below 10GeV, is approximately consistent with the expectation values being entirely secondary. If additional sources other than secondary positrons existed, there is the attractive problem of the production and accelerations of cosmic rays. Ginzburg [28] and Hayakawa et al. first pointed out the importance of measuring the fraction of the flux of Positrons to Electrons in 1958 [28], [24].

The flux of electron is small, under 1% of the overall cosmic ray flux beyond a few GeV, and for precise measurements we need to identify the electrons by rejecting the much more abundant hadronic showers seen in detectors. That is the reason why the first measurements of primary electrons were delayed until 1961, compared to other components of cosmic rays.

The first successful observation was made by P.Meyer and R.Vogt (1961) with scintillation counters and by J. Earl (1961) with an imaging detector, a balloon borne cloud chamber. Many experiments were performed since the discovery of electrons in 1960s. Among the many works on the observation of electrons, I show an example of the observations made by Japan and US collaborations around 1980 [29].

The detector is the emulsion chamber, which is a sandwich of lead plates and nuclear emulsions coated on both sides of a thin plastic plate as shown in Fig. 13. Electrons are identified by tracing showers back to the primary electron track using a microscope. Electron showers begin with an associated electron pair created by the primary electrons within a top layer of a few radiation lengths of the detector. The rejection power to proton is estimated as $10^4$-$10^5$. The detailed will be found in the original paper [29]. The energies of electrons are identified by counting the number of shower tracks within the 100mirons from then shower axis. As illustrated in the Fig.14, we see, no particular structure on the electron spectrum was observed beyond the statistical errors.

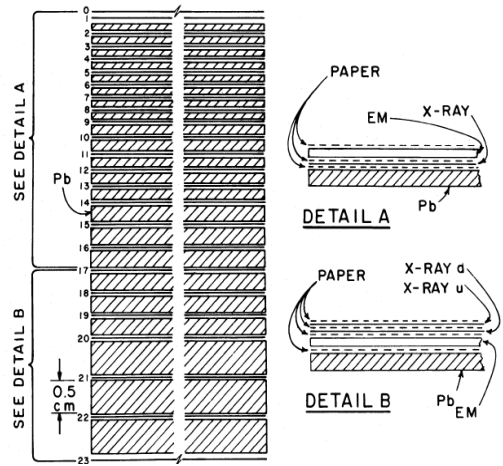

Figure 13 : Emulsion chamber configuration as a Detector of cosmic-ray electrons in 1976 flight [29].

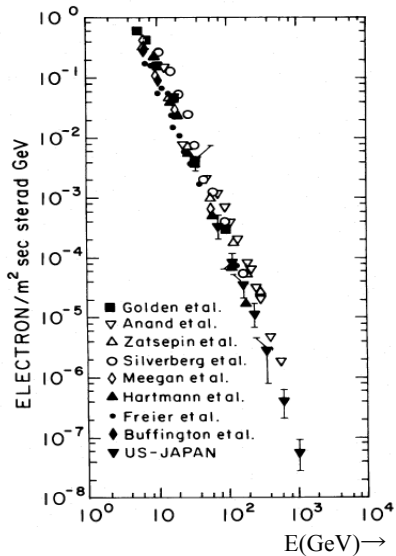

Figure 14. Observed Cosmic-Ray electron Spectrum around 1980 [28]

A large total exposure factor was accumulated by a series of balloon observations, by 2000 almost seven times of those of [28] of 1980 (finally ~8.2m²sr day beyond 1TeV), and electrons up to a few TeV region were observed in these detectors [29].

### 4.2.1. Effects of nearby sources of Electrons

The theoretical argument on the possible deviation of the smooth power law of electron spectrum was first mentioned by C.S.Shen [30] based on the Pulsar and Supernova origin of cosmic rays in 1970.

The electrons lose energy almost proportional to the square of their energies, by the Synchrotron and Inverse Compton processes. Then if electrons of energy of E are observed at the Earth, they must be emitted from a source within the past T years, where T is inversely proportional to the energy of E. The value of T also depends on the energy densities of ambient photons and magnetic field. As an illustration, using the proper energy density of ambient photon and magnetic field, we estimate that for electrons with E>1TeV must have been produced within T<10⁵ years.

During this lifetime of T, 1TeV electron can travel about 1kpc depending on the values of diffusion parameter. If the energy is smaller than 1TeV, the lifetime T is longer, and the travel distance increases. This means if we look the higher energy spectrum of electrons they must have been produced more recently than those of low energy electrons. Accordingly, the distance of their sources must be nearer. As the energy of electrons become higher, the location of the source must be nearer, and the number of available sources (SNR and Pulsars) is limited. In the higher energy region, we expect only a few sources for electrons, and we would expect large non-statistical fluctuations of the electron spectrum and anisotropy for nearby sources.

Each individual source might create a feature in the spectrum. We would therefore expect to observe humps and the anisotropies in the spectrum, corresponding to the identifiable sources.

These describe the concept by Shen, and more details will be found in his paper [31]. When Shen proposed this concept, he assumed sources were the observed SNR and pulsars, but the parameters of those objects were not clear at the time. Later, several authors, Cowsik-Lee (1979), Nishimura et al. (1979), Aharonian et al. (1995), Atoyan et al. (1995), Pohl-Espoid (1998), Erykin–Wolendale (1998) and Kobayashi et al. (2004), discussed these features more details using the most recent data of those objects [32], [33]. I presented some of the results in early days to the international Conference of Cosmic rays in Kyoto, 1979.

E³×Flux ( electrons /m² s.sr. GeV⁻²)

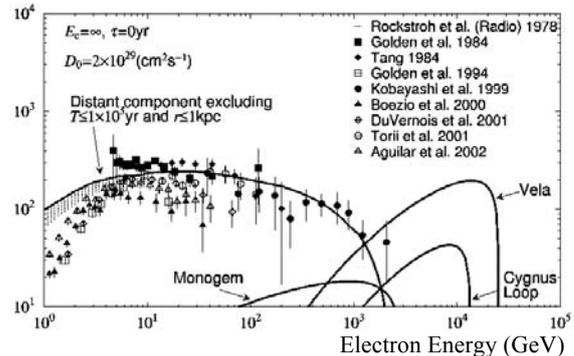

Figure 15: An example of the effect to Cosmic-ray electron spectrum by nearby sources [33]

However, the rapporteur of my friend, did not refer at all this work. He explained that such event might occur at extremely high-energy region where the flux is few, then it could not be observed and the argument is not realistic and he said to me why he discarded my report.

Such response is some times occur when the new concept were proposed. After 30 years from this episode, the hump of electron spectrum becomes one of the important phenomena relating to the origin of cosmic rays and even to the existence of the Dark Matters, which are now to be discussed in this meeting.

### 4.2.2. Observed hump in the Electron Spectrum

The hump of electron spectrum was first reported in the series of the observations of large balloon-borne ATIC detector program.

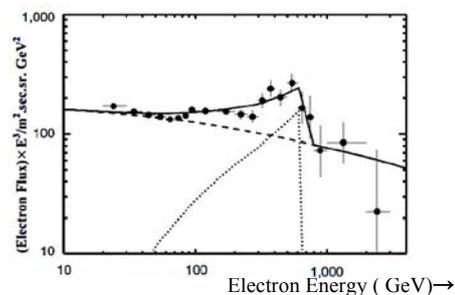

Figure 16: Observed hump in the electron spectrum between 300-800GeV by ATIC group [34].

J. Chang et al. of ATIC group claimed the excess of cosmic-ray electrons at energies of 300GeV – 800 GeV. which could be interpreted as due to the nearby source of electrons or due to the pair electrons from the annihilation of dark maters of mass of around 600GeV [33]. Their data are shown in Fig. 16.

Several observations followed to provide more details of the electron and positron spectrum relating to these indications.

HESS presented the spectrum of primary electrons observed through Cherenkov radiation from the extensive air showers of the primary electrons [35]. This indicates the decline of the spectrum beyond 1 TeV.

### 4.2.3. Positron Excess

The Pamela Satellite, a magnet spectrometer detector which was launched in 2006, found a definite increase of the positron fraction from 10 to 100GeV, which indicated the existence of positron sources other than secondary positrons from muon decay [36].

The Fermi satellite observed the primary electron spectrum and also estimate the positron fraction by exploiting the East and West Asymmetry of the electron components [37]. Most recent data are due to AMS (Alpha Magnet Spectrometer), which has almost ten times larger acceptance area of Magnetic spectrometer than Pamela. AMS was launched and installed on International Station in 2011, and observed more significant data than Pamela on the positron fraction as well as electron spectrums [38].

A summary of the data on Cosmic-ray electrons from these recent observations is shown in Fig. 19. The hump at several hundred GeV exists, but looks to be smeared in shape by combining those data compared to the hump seen by ATIC. The fraction of positron is definitely increased up to 500GeV, indicating the existence of sources other than secondary production from the decay of muons, but increasing rate ceases beyond 200GeV as shown in Figure 20

Then the problems are what are the sources of high-energy electrons and particularly the positrons.

PAMELA
**P**ayload for **A**nti**m**atter **E**xploration and **L**ight **N**uclei **A**strophysics

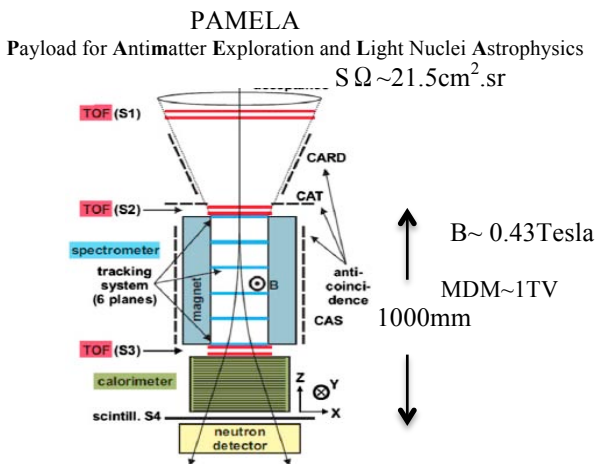

Figure 17: Configuration of Pamela,
Margent spectrometer for cosmic rays. In orbit 2008 [36].

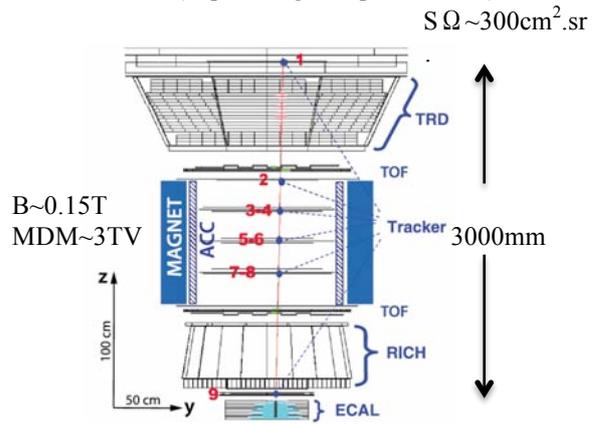

Figure.18: AMS (Alpha magnet Spectrometer)
Large size spectrometer to observe High Energy Cosmic Rays, installed on International station in 2011. [38]

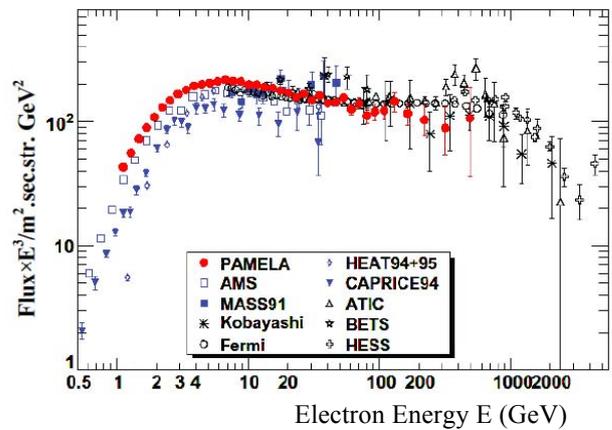

Figure. 19: Electron Spectrum; from [36]
Most recent data of AMS [38] are not included, but the AMS data is limited to ~500GeV. Below a few hundred GeV the data of AMS are consistent with those of Pamela [36].

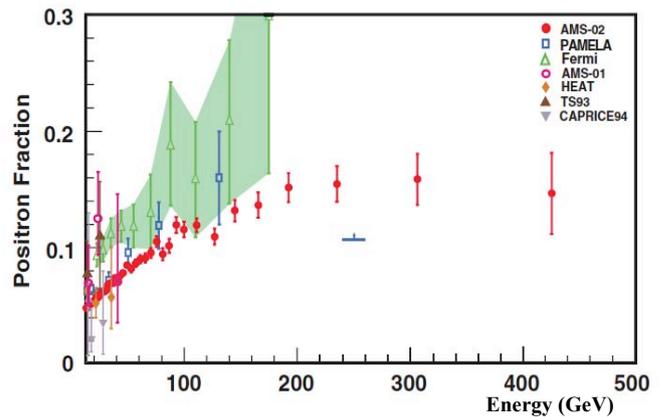

Figure.20: Positron Excess
The fraction of positron is definitely increasing with positron energy. This indicates the existence of sources other than secondary product of muon decays, but increasing rate ceases beyond 200GeV [38]. The fraction of the secondary positrons from muons is estimated under 0.02 beyond 100GeV [36].

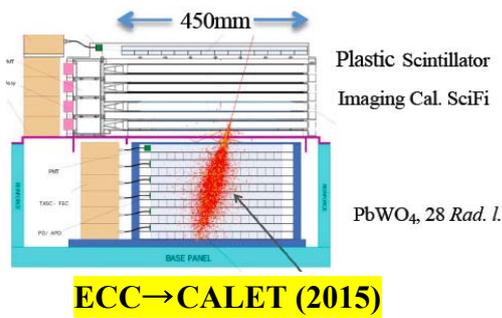

**ECC→CALET (2015)**

Figure 21:  Calet detector layout
 (**Cal**orimetric **E**lectron **T**elescope) [39]
Inside bracket shows the expected launching yr.

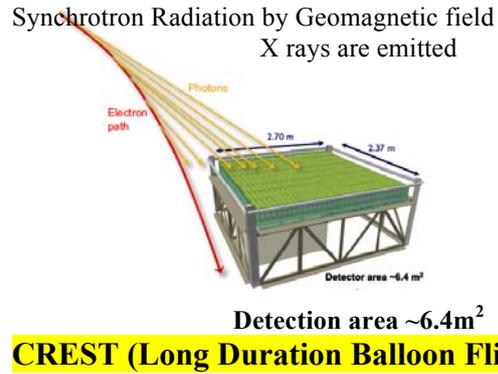

Synchrotron Radiation by Geomagnetic field
X rays are emitted

**Detection area ~6.4m²**

**CREST (Long Duration Balloon Flights )**

Figure 23: CREST detector Layout
(**C**osmic-**r**ay **E**lectron **S**ynchrotron **T**elescope) [41]

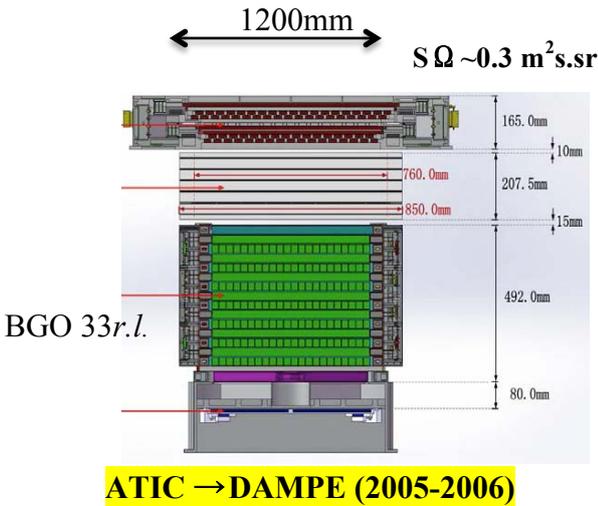

**ATIC →DAMPE (2005-2006)**

Figure 22:  DAMP detector layout [40]
     (**Da**rk **M**atter **P**article **E**xplore)
Inside bracket shows the expected launching yr.

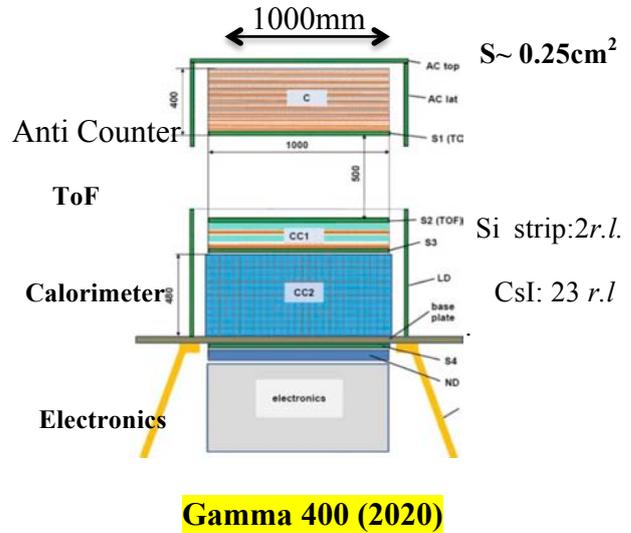

**Gamma 400 (2020)**

Figure 25:  Gamma 400 detector layout [42]
Inside bracket shows the expected launching yr.

   Possibilities are discussed of nearby sources by pulsars and/or SNR surrounded by  gas material and /or dark matter. In case of Dark matter annihilation, positron fractions increase with positron energies, but should drop abruptly beyond the rest energy of Dark matter. Similar but not so sharp feature would also be seen in case of nearby source. These ideas are to be discussed in this meeting. To obtain a more precise spectrum of electrons from the GeV to TeV region, several new programs are planned to be in operation within a few years. Those are shown in Fig. 21, 22, 23 and 24.  The detectors have essentially large detection area with deep depth of calorimeter or to detect the Geo-Synchrotron X-rays together with the detection of the arrival direction of incoming particles to see the possible anisotropies of the particles. These are enable us to detect small flux of electrons and gamma rays with high rejection power against to the hadronic components. These experiments will bring us new findings relating the sources of high-energy electrons and /or related dark matter in near future.

## 5. Summary and Acknowledgements

   In describing the "Dawn of high energy Astrophysics Japan", I found we owe very much to Hayakawa for his tremendous efforts to develop this field with his pioneering works and stimulation in our country. We deeply appreciate him for his outstanding leadership for many years from his young days in early 1950. I hope the success of this symposium in Nagoya through good discussions, explorations and new findings.
   In closing my talk, I would like to acknowledge to the organizing committee for inviting me to talk this subject. I also wish to thank to my colleagues for their useful comments in preparation of this manuscript.